\documentstyle[12pt]{article}
\begin{document}
\textwidth=16cm
\textheight=21cm

\title
{{\rightline{\small{TIFR-TH/95-38, MIT-CTP-2459}}}
{\rightline{}}
Folds, Bosonization and  non-triviality of the classical limit of
2D string theory}
\author{Sumit R. Das \thanks{E-mail: das@theory.tifr.res.in}\\
{\small{{\em Tata Institute of Fundamental Research}}} \\
{\small{{\em Homi Bhabha Road , Bombay 400005, INDIA.}}}
\and
Samir D. Mathur\thanks{E-mail: me@ctpdown.mit.edu} \\
{\small {\em{Center for Theoretical Physics,
Massachussetts Institute of Technology}}}\\
{\small{\em{Cambridge, MA 02139. U.S.A.}}}}
\date{}
\maketitle

\begin{abstract}
In the 1-dimensional matrix model one identifies the tachyon field in
the asymptotic region with a nonlocal transfom of the density of
fermions. But there is a problem in relating the classical tachyon
field with the surface profile of the fermi fluid if a fold forms in
the fermi surface. Besides the collective field additional variables
$w_j(x)$ are required to desrcibe folds. In the quantum theory we show that
the $w_j$ are the quantum dispersions of the collective field.
These
dispersions become $O(1)$ rather than $O(\hbar)$ precisely after fold
formation, thus giving additional `classical' quantities and leading
to a rather nontrivial classical limit. A coherent pulse reflecting
from the potential wall turns into high energy incoherent quanta (if a
fold forms), the frequency amplification being of the order of the square
root of the number of quanta in the incident wave.
\end{abstract}
\newpage

The double scaling limit of the one dimensional matrix model provides
a non-perturbative formulation of two dimensional string theory in the
Liouville background \cite{CEQONE}.
Classically, in the two dimensional string theory
one expects that a matter
wave with sufficient energy and energy density will collapse
and a black hole will be formed. The matrix model should describe
this classical process. However, since the graviton and dilaton
are not present as explicit fields in the matrix model, collapse
would be signalled by properties of the matter alone, e.g. an
infinitely long time delay for return of matter thrown into
the liouville wall.
In addition, the matrix model should be able to
answer questions in black hole physics at the quantum level
{\em exactly}.

However, it turns out to be surprisingly difficult to understand a
potential gravitational collapse in the matrix model.
The natural field which describes
the matrix model scalar - the collective field - is {\em not} the
string theory massless tachyon. Recent work has revealed that for
{\em weak} waves of low energy the string theory tachyon may be
expressed, in the asymptotic region,
as a {\em non-local} transform of the collective
field \cite{POLTRANS}
which is the fourier transform of the well-known ``leg-pole''
factor \cite{LEGPOLE}.
The nonlocal transform then relates the space of
eigenvalues in the matrix model with the space in string theory.

Unfortunately we cannot use this correspondence for waves of {\em
finite} amplitude and width, in particular for the ones which we expect will
form
black holes.
 At the fundamental level the
matrix model is equivalent to a system of $N$
mutually  noninteracting fermions
in a background inverted harmonic oscillator
potential.
The classical state of the system is specified by a distribution
function in phase space $u(x,p,t)$ - in fact $u(x,p,t) = {1 \over
2\pi \hbar}$ in some region bounded by the fermi surface.
We will consider only
states which correspond to a single connected filled region of
phase space. We will find it convenient to parametrize the fermi
sea in terms of an infinite number of functions $w_n (x,t)~~(n =
0,1,\cdots)$ following \cite{AVJEV}
\begin{eqnarray}
\int dp~u(x,p,t) & = & {1 \over 2\pi \hbar}[\beta_+ - \beta_-] \nonumber \\
\int dp~p~u(x,p,t) & = & {1 \over 2\pi \hbar}[{1\over 2}(\beta_+^2
- \beta_-^2) + (w_{+1}-w_{-1})] \nonumber \\
\int dp~p^2~u(x,p,t) & = & {1 \over 2\pi \hbar}[{1\over 3}(\beta_+^3
- \beta_-^3) \nonumber \\
& & + (\beta_+w_{+1} - \beta_-w_{-1}) + (w_{+2}-w_{-2})]
\label{eq:nine}
\end{eqnarray}
and so on. Let a line of constant $x$ in  phase space
cut the upper edge of the fermi sea at points $p_i(x,t)$
and the lower edges of the fermi sea at $q_i(x,t)$ where
$ (i = 1,2,\cdots i_m(x,t))$. Then
\begin{equation}
\int dp~p^n~u(x,p,t) = {1 \over 2\pi \hbar(n+1)}\sum_{i=1}^{i_m}
[p_i^{n+1} - q_i^{n+1}]
\label{eq:eleven}
\end{equation}
$i_m(x,t)- 1$ denotes the number of ``folds'' at
the point $x$ at time $t$. In the absence of folds, one can
set $\beta_+(x,t) = p_1 (x,t)$ and
$\beta_-(x,t) = q_1 (x,t)$ which then implies that all the
$w_{\pm , n} = 0$. This is the standard bosonization in terms of
the collective field theory. As emphasized in \cite{POLF}
and \cite{WADIAF}
only in the absence of folds  is the state  described by
a scalar field $\eta(x,t)$ and its momentum conjugate $\Pi_\eta
(x,t)$,  which are related to $\beta_\pm$ by $\beta_\pm
= \Pi_\eta \pm \partial_x \eta$.
In the asymptotic region the string theory
massless tachyon is a nonlocal transform of $\eta (x,t)$.
In the presence of folds we do not know how to extract the string
theory space-time from the matrix model, since the $w_{\pm,n} \ne
0$ and are needed to specify the classical state of the fermi fluid while in
the
far asymptotic region the collective field configurations
alone exhaust the possible configurations of the string theory
tachyon field
\footnote{Away from the
asymptotic region higher moments of $u(x,p,t)$ are required for
determining the string theory tachyon. See Dhar et. al. in
\cite{POLTRANS}.}.

Under time evolution folds will generically form even if we start with
profiles without any fold.  Nevertheless, if we start with a
non-folded tachyon wave with sufficiently small amplitude (for a given
wavelength) the outgoing pulse at late times does not develop a fold.
This is why one can specify the incoming and outgoing states by the
collective field and obtain an $S$- matrix by performing a
perturbation expansion in the strength of the wave. However, for a
pulse which is sufficiently narrow and/or tall the outgoing pulse
in the asymptotic region can
have a fold even if the ingoing one did not.  An extreme example is a
pulse which is tall enough that its tip goes over to the other side of
the potential barrier - it is these pulses for which we may expect
black holes to form. We wish to emphasize that fold formation is
invisible if the classical collective field equation is solved
{\em only} in a perturbation expansion (in the weakness of the field).
In this sense it is a genuinely nonlinear and nonperturbative effect.

In this paper we give the physical meaning of these ``folds''.  (We do
not however address the implications of fermi fluid crossing the
potential barrier.)  A basic question that arises is: under the
boson-fermion map what is the bose state corresponding to the
classical fermi fluid with fold? (The choice of map is distinct from
the choice of Hamiltonian. In particular we can ask for the
bosonisation of fold states for relativistic fermions as well.)

We will show that a folded configuration correspond to a rather
nontrivial classical limit. Normally the classical limit of a quantum
system is obtained by considering coherent states.  The quantum
fluctuations of Heisenberg picture operators in such states are
suppressed by powers of $\hbar$ so that their expectation values
become classical dynamical variables in the $\hbar \rightarrow 0$
limit.  We will argue that for fermionic systems under consideration,
folded classical configurations correspond to non-coherent states
which have a large $O(1)$ dispersion.

For relativistic fermions
a (Schrodinger picture)
coherent state cannot develop into such a noncoherent state under
time evolution. However, for nonrelativistic fermions this happens
generically in a time interval of order unity. In fact the
time at which this happens is exactly equal to the time at
which a fold forms in the classical picture.
We will demostrate this explicitly in a model closely
related to the fermionic field theory of the matrix model. The derivation
makes it clear that the physics is entirely similar in the matrix
model. This allows a natural way to identify the space-time in
the matrix model based on a map of the quantum operators.

Let us first discuss
the dynamics of fold formation in the classical
theory. As shown in \cite{AVJEV} the standard classical fermion algebra
is consistent with imposing the Poisson brackets
\begin{eqnarray}
[\beta_\pm (x,t),\beta_\pm (y,t)]_{PB} & = & 2\pi\hbar\partial_x \delta (x-y)
\nonumber \\
\left[w_{\pm , m}(x,t), w_{\pm , n}(y,t)\right]_{PB} &
= & 2\pi\hbar[m w_{\pm ,m+n-1}(x,t) \nonumber \\
& & + n w_{\pm ,m+n-1}(y,t)]\partial_x \delta (x-y) \nonumber \\
\left[w_{\pm ,n}(x,t),\beta_{\pm}(y,t) \right]_{PB} & = & 0
\label{eq:ten}
\end{eqnarray}
It is well known that powers of $\beta_{\pm}$ satisfy a $w_{\infty}$
algebra. In this construction the $w_{\pm,n}$ themselves satisfy
an independent $w_\infty$ algebra. \footnote{In \cite{AVJEV} it
was also found that these statements are true at the quantum level
as well ($W_\infty$ algebra) and the central charge of
$w_{\pm,n}$ turns out to be zero. In this paper we will not be
concerned with the quantum algebra.}

So far our considerations were independent of the
fermion dynamics. Given the hamiltonian we can now obtain
the time evolution equations for the quantities $\beta_\pm,
w_{\pm,n}$ using the above definitions and the Poisson brackets.
Since the $\pm$ components are decoupled we can treat only one
of them. In the following we will omit the $\pm$ subscript.
We will also concentrate on {\em free} fermions.
A background potential can be easily incorporated and is not
essential to the main point. For relativistic fermions one has
a hamiltonian $H = \int dx \int dp~p~u(x,p,t)$. Using (\ref
{eq:nine}) and (\ref{eq:ten}) we easily get
\begin{equation}
\partial_t\beta = -\partial_x \beta~~~~~\partial_t w_m
= -\partial_x w_m
\label{eq:fifteen}
\end{equation}
For nonrelativistic fermions one has $H = \int dx
\int dp~{1\over 2}p^2~u(x,p,t)$
and using the Poisson brackets one gets the evolution equations
\begin{equation}
\partial_t \beta  =  -\beta\partial_x\beta -  \partial_x w_1
;~~~\partial_t w_m  =  -2w_m~\partial_x\beta - \beta~
\partial_x w_m - \partial_x w_{m+1}
\label{eq:seventeen}
\end{equation}

The $\beta, w_n$ are expressible in terms of the $(p_i,q_i)$
introduced in (\ref{eq:eleven}). We choose a parametrization
associated to the upper edge of the fermi surface, i.e. we will choose
all the $w_{-,n} = 0$ and
\begin{equation}
\beta_-(x,t) = q_{i_m}~~~~\beta_+ (x,t) = \sum_{i=1}^{i_m} p_i
- \sum_{i=1}^{i_m-1} q_i
\label{eq:twelve}
\end{equation}
This identification is natural when the fold is close
to the upper edge of the fermi sea but may be used even
for a fold near the lower edge.
We can then use the same identification
for relativistic fermions where the fermi sea has no lower
edge and one has to set all $\beta_- = w_{-n} = 0$.
In fact since the question of folds has nothing to do
with the presence or absence of a lower edge of the fermi sea
we will simply imagine that the fermi sea is bottomless even
for nonrelativistic fermions. This is a good approximation for boson energy
densities small enough to not reach the bottom of the fermi sea in the fermi
picture.
Using (\ref{eq:twelve}), (\ref{eq:nine}) and (\ref{eq:eleven})
we can now easily calculate all the $w_{\pm,n}$ 's. Since the
$p_i,q_i$ do not contain any $\hbar$, the $w_n$ 's, if nonzero,
are also independent of $\hbar$.

Each of the
$p_i, q_i$ satisfy an evolution equation determined by the single
particle hamiltonian: $\partial_t p_i = - \partial_x p_i$ for
relatvisitic fermions and $\partial_t p_i = -p_i\partial_x p_i$
for nonrelativistic fermions (and similarly for $q_i$'s). It
may be checked that these equations then imply the corresponding
evolutions for the $\beta$ and $w_m$ 's.

Let the
profile of the fermi surface at $t = 0$ be given by
$p(x,0) = a(x)$. Then at a later time $t$ it is easy to show that
\begin{eqnarray}
p(x,t) = a(x-t) &~~~~~&{\rm relativistic}
\nonumber \\
p(x,t) = a(x-p(x,t)~t) &~~~~~&{\rm nonrelativistic}
\label{eq:nineteen}
\end{eqnarray}
If there is a fold, there must be some point where
${dx \over dp} = 0$ \cite{MENDE}.
Since the profile in a relativistic system
is unchanged in time a fold cannot develop from a profile
with no folds. On the other hand for a nonrelativistic system,
even if we start with a single valued  $a(x)$
the solution to the second equation in
(\ref{eq:nineteen}) will generically lead to ${dx \over dp} =0$
at some point on the fermi surface at some time.
We give the result for the time of fold formation $t_f$ for two
initial profiles:
\begin{eqnarray}
p(x,0) & = & b~e^{-{(x-a)^2 \over c^2}}- b~e^{-{(x+a)^2 \over c^2}},~~~~~~
t_f \approx {c e^{{1\over2}}\over b{\sqrt{2}}}~~~~
 {\rm for} ~~{a\over c}>>1  \\
p(x,0) & = & k~{\rm Re}(C_k~e^{ikx}),~~~~~~
t_f = {1 \over (\vert C_k \vert k^2)}
\label{eq:twenty}
\end{eqnarray}
These examples clearly show that fold formation can occur for
pulses of arbitrarily small energy density (and total energy),
provided the  width is sufficiently small.

It is now clear what happens if we set $w_n = 0$ in the time
evolution equations (\ref{eq:seventeen}). The evolution of $\beta$
proceeds smoothly till $t = t_f$ at which point $\partial_x \beta$
diverges. The equations cannot be evolved beyond this
point and the collective field theory fails.
The time evolution of $u (p,q,t)$ is of course unambiguous
\cite{WADIAF}. The point is that one needs nonzero
values of $w_n$ beyond this time, whose presence
render the classical
time evolution of the entire $\beta,w_n$
system completely well defined.
While we have illustrated this
phenomenon in the simple case of free nonrelativistic fermions, it
is clear that the conclusions are same for the inverted harmonic
oscillator potential which appears in 2d string theory.

We now turn to the quantum theory.
Consider $N$ free fermions of mass $M = 1$ in a box
of size $L$. Define the field operators
\begin{equation}
{\hat \psi}(x) = {1\over \sqrt{L}}\sum_{n=-\infty}^{\infty}
{\hat \psi}_n~e^{i{2\pi
n \over L}x} ~~~~~~~~~~~~~
{\hat \psi^\dagger}(x) = {1\over \sqrt{L}}\sum_{n=-\infty}^{\infty} {\hat \psi
^\dagger}_n~e^{-i{2\pi n \over L}x}
\label{eq:five}
\end{equation}
where ${\hat \psi}_n$, ${\hat \psi
^\dagger}_n$
are annihilation and creation operators for a fermion at level $n$.
One has the standard anticommutation relations
\begin{equation}
[{\hat \psi_n}^\dagger,{\hat \psi_m}]_+ ~=~\delta_{n,m},~~
[{\hat \psi_n},{\hat \psi_m}]_+~=~[ {\hat \psi_n}^\dagger,{\hat
\psi_m}^\dagger]_+~=~0.
\label{eq:six}
\end{equation}
As mentioned above for the purpose of discussing the physics
of folds the presence of a lower edge of the fermi sea for
nonrelativistic fermions is an unnecessary complication.
Henceforth we
will assume that the fermi sea is bottomless.
Thus the vacuum is defined, for both relativistic and nonrelativistic
fermions by \footnote{We have chosen phases in (\ref{eq:five})
so that $n = 0$ is the vacuum fermi level}
\begin{equation}
{\hat \psi}_n |0> = 0 ~~{\rm for}~~n > 0 ; ~~~~~~~~~~
{\hat \psi}^\dagger_n |0> = 0~~{\rm for}~~ n \le 0
\label{eq:eight}
\end{equation}

Let
\begin{eqnarray}
&\alpha_{-n}(t)~=~\sum_{m=-\infty}^\infty :{\hat\psi}^\dagger_{n+m} (t)
 \hat\psi_m (t):
{}~~~~{\rm or}~~~~\alpha (x,t) = :{\hat\psi}^\dagger(x,t)\hat\psi(x,t):
\nonumber  \\
&\alpha(x,t)~ =
{}~{1\over {L}}\sum_{n=-\infty}^\infty \alpha_n (t)
e^{i{2\pi\over L} nx}~\equiv ~\partial_x\phi (x,t)
\label{eq:twentytwo}
\end{eqnarray}
The modes $\alpha_n (t)$ are thus shift operators on fermion levels.
(See e.g. \cite{KACRAINA}.) This bosonisation is
exact as long as the action of the $\alpha_n$ do not move a
state past the bottom of
the fermi sea; in particular it is exact if the sea is bottomless.
The normal ordering is defined as
\begin{equation}
:{\hat \psi_{m}}^\dagger \hat \psi_n:~=~{\hat \psi_{m}}^\dagger \hat \psi_n
{}~~{\rm if}~ n>0~~~~~~~~
:{\hat \psi_{m}}^\dagger \hat \psi_n:~
=~- \hat \psi_n {\hat \psi_{m}}^\dagger~~{\rm if}~ n\le 0
\label{eq:twentythree}
\end{equation}
Then it may be verified that
\begin{equation}
[\alpha_n (t),\alpha_m (t)]~=~n\delta_{n+m,0}~~~~{\rm or}~~~~~
[\alpha(x,t),\alpha(x,t)]~=~-i{1\over 2\pi} \delta'(x-x')
\label{eq:twentyfour}
\end{equation}
The inverse correspondence to (\ref{eq:twentytwo}) is given as follows:
\begin{eqnarray}
\hat \psi(x) & = & {1\over \sqrt{L}}e^{-i\beta_0}~e^{-i{2\pi\over L}x\alpha_0}~
e^{-\sum_{j\ge 1}{\alpha_{-j}\over j}~e^{i{2\pi\over L} jx}}  e^{\sum_{j\ge
1}{\alpha_{j}\over j}~e^{-i{2\pi\over L} jx}} \nonumber \\
{\hat \psi(x)}^\dagger
& = &{1\over \sqrt{L}}e^{i{2\pi\over L}x\alpha_0}~e^{i\beta_0}
 ~e^{\sum_{j\ge 1}
{\alpha_{-j}\over j}e^{i{2\pi\over L}
jx}} ~ e^{-\sum_{j\ge 1}{\alpha_{j}\over j}e^{-i{2\pi\over L} jx}}
\label{eq:twentyfive}
\end{eqnarray}
where the zero-mode operators satisfy $[\alpha_0,\beta_0]=-i$.
In the classical limit, the classical quantities in (\ref{eq:nine})
are then related to expectation values of appropriate
operators in some quantum state, e.g.
\begin{eqnarray}
\int dp~u & = & < :\psi^\dagger\psi:> = <\alpha (x,t)>
\nonumber \\
\int dp~p~u  & = & <{i \hbar \over 2}
[ : (\partial_x \psi^\dagger)\psi
-\psi^\dagger (\partial_x \psi ):]> = 2\pi\hbar <:
{\alpha^2\over 2}:> \nonumber \\
\int dp~p^2~u  & = & < {-\hbar^2 \over 6}
:[\psi^\dagger (\partial_x^2\psi) + (\partial_x^2\psi^\dagger)\psi
- 4(\partial_x\psi^\dagger)(\partial_x\psi)]: \nonumber \\
& = & 2\pi^2\hbar^2
<:{\alpha^3\over 3}:>
\label{eq:twentyeight}
\end{eqnarray}
and so on. The second equalities in (\ref{eq:twentyeight}) follow
from the operator correspondence (\ref{eq:twentyfive}) and may be
efficiently obtained using operator product expansions.
The equations (\ref{eq:twentyeight}) demonstrate the main point
of this paper : {\em $w_n$ are a measure
of the quantum fluctuations in the bosonic field}. Thus, e.g.
comparing (\ref{eq:nine}) and (\ref{eq:twentyeight}) we have
\begin{equation}
w_1 (x,t) = (2\pi\hbar)^2 [{1\over2}<:\alpha^2(x,t):> -
{1\over2}<\alpha (x,t)>^2]
\label{eq:twentynine}
\end{equation}
Note that in the normalizations we are using,
the fluctuations of $\alpha$ must be
$O({1 \over \hbar^2})$ for $w_n \sim O(1)$.

We now consider the evaluation of $w_1(x,t)$ in a Heisenberg
picture coherent state of bosons
\begin{equation}
| \psi > = \prod_{n=1}^{\infty} e^{{C_n \over \hbar} \alpha_{-n}(0)} | 0 >
\label{eq:thirty}
\end{equation}
It is straightforward to check that
\begin{equation}
< \alpha (x,0) > \equiv {<\psi | \alpha (x,0) | \psi> \over
< \psi | \psi>} = {2 \over \hbar L}\sum_{n=1}^\infty
{\rm Re}[n~C_n~e^{2\pi~i~n~x/L}]
\label{eq:thirtyone}
\end{equation}
Note we have chosen our state such that $<\alpha (x,0)> \sim
{1\over \hbar}$ which is required for $\beta \sim O(1)$
(see  (\ref{eq:twentyeight})). It may be easily shown that
\begin{equation}
<:\alpha^n (x,0):> = <\alpha (x,0)>^n
\label{eq:thirtytwo}
\end{equation}
Thus the $w_n$ vanish at $t=0$, and we have a fermi surface without folds.
In the following we will consider a state with only one nonzero
$C_n$ for $n = {\bar n}$. The classical fluid profile is then
given by (\ref{eq:twenty})
This is sufficient for our purpose.

In the Heisenberg picture, the fermion operators evolve as
\begin{equation}
\hat\psi_n(t)~=~\hat\psi_n e^{-i({2\pi\over L})^2 {\hbar
\over 2} n^2 t}
{}~~~~~{\hat\psi_n}^\dagger(t)~=~{\hat\psi_n}^\dagger
e^{i({2\pi\over L})^2 {\hbar \over 2} n^2 t}
\label{eq:thirtyfive}
\end{equation}
The modes of the bosonic field then evolve as
\begin{equation}
\alpha_n (t) =  e^{i{\hbar\over 2}({2\pi\over L})^2 n^2 t}
\sum_m:{\hat\psi_{-n+m}}^\dagger \hat\psi_m: \lambda^m(n)
\label{eq:thirtysix}
\end{equation}
where
\begin{equation}
\lambda(n) = exp[-i\hbar ({2\pi \over L})^2~n~t]
\label{eq:threesix}
\end{equation}
Using the evolution of the fermion operators we can
compute the expectation value of the bose operators.
The details of the calculation will be given in \cite{DASMATHUR2};
here we note the results.  Define
\begin{equation}
\Phi(n,p) = {4 |C_n| \over
\hbar}~{\rm sin}~[({2\pi \over L})^2~{\hbar p n t \over 2}]
\label{eq:fortyone}
\end{equation}
Let $p/{\bar n}$ be an integer. Then the one point function is
\begin{equation}
<\alpha_p(t)>~=
{}~(-i)^{p/{\bar n}}
{\lambda(p)\over \lambda(p)-1}
({C^{1\over 2}_{\bar n}\over {C^{1\over 2}_{\bar n}}^*})^{{p\over \bar n}}~
J_{p/{\bar n}} (\Phi({\bar n},p))
\label{newone}
\end{equation}
($J_n(z)$ is the Bessel function.) If $p/{\bar n}$ is not an
integer then $<\alpha_p(t)>=0$.
Similarily, the two point function is nonzero only if
$(p+q)/\bar n$  an integer:
\begin{eqnarray}
<\alpha_p(t)\alpha_q(t)>_c~ & \equiv~ &<\alpha_p(t)\alpha_q(t)>
- <\alpha_p(t)><\alpha_q (t)> \nonumber \\
& = &
\sum_{s > {q \over {\bar n}}} (-i)^{p+q\over \bar n}
({C^{1\over 2}_{\bar n} \over {C^{1\over 2}_{\bar n}}^*})^{p+q \over {\bar n}}
{}~~J_s(\Phi({\bar n},q))~J_{{p+q \over {\bar n}}-s}(\Phi ({\bar n},p))
\nonumber \\
& & \lambda(p) \lambda (q) {(\lambda(p)\lambda(q))^
{(q-{\bar n}s)/2}-(\lambda(p)\lambda(q))^{
{(\bar n}s - q)/2}\over 1 - \lambda(p)\lambda(q)}
\label{eq:fortytwo}
\end{eqnarray}
As a special case we note $G(p,t) \equiv <\alpha_{-p}(t)
\alpha_p(t)>_c$
\begin{equation}
G(p,t) = \sum_{s > {p \over {\bar n}}}
({\bar n}s - p)
J_{s}^2(\Phi({\bar n},p))
\label{eq:fortythree}
\end{equation}

One may verify that the result (\ref{eq:fortythree}) reproduces  at $t = 0$
the expected
result that $<\alpha_q(t)\alpha_p(t)>_c = q\theta(q)~\delta_{p+q,0}$.

We can estimate on heuristic grounds the quantity $G(p,t)$ by
considering the Schrodinger picture of evolution. The state $|\psi(t)>$
describes the fermi fluid, with a fold after some time $t_0$.
Let $|p/\bar n|>>1$, $p>0$.
For the action of such $\alpha(p)$ we may consider a succesion of small $x$
space
intervals, over each of which the sections of the fermi surface are
approximately constant at heights $p_i,q_i$.  The operator $\alpha_p$ is a
lowering operator,
which destroys a fermion at some level $m$ and creates one
at a level $m - p$. If there is no fold then $G(p,t)=0$ since fermions cannot
be lowered further into the already occupied levels. If there is one fold then
$\alpha_p$ can move fermions from the interval $(p_1,q_1)$ to the vacant levels
in $(q_1,p_2)$. The operator $\alpha_{-p}$ moves the fermions back, and we find
$G(p,t)\ne 0$. More generally we get a result proportional to the number of
vacant `bands' which equals the number of folds.
It may be also seen that $G(p) = 0$ for $p$ very large. This is because
for very large $p$, $\alpha_p$ must lower a fermion to a level inside the main
bulk
of the fermi sea. Since the {\em momenta} in the fermion picture are
all of order unity, $G(p)$ should valish rapidly for $p > p_M$ where
$p_M \sim O({1\over \hbar})$.

The result (\ref{eq:fortythree}) has exactly this feature.
First, it may be seen from (\ref{eq:fortyone})
that when
\begin{equation}
p > p_M = {4 |C_{\bar n}|~{\bar n} \over \hbar}
\label{eq:fournine}
\end{equation}
$G(p)$ vanishes exponentially with increasing $p$ regardless of
the $\hbar \rightarrow 0$ limit. This is because in this case
the index of the Bessel function is {\em always} greater than the
argument and Bessel functions decay exponentially when the ratio of
the index to the argument grows large (see (\ref{eq:fortyfour}) below).

Consider now the classical limit for $p >> {\bar n}$, but $p < p_M$.
In the $\hbar \rightarrow 0$ limit one has
has $\lambda (p) = - \Psi({\bar n},p) = 1$ and $\Phi({\bar n},p) =
2 \vert C_{{\bar n}} \vert ({2\pi \over L})^2~{\bar n}~p~t$.
Because of the exponential decay of the Bessel function,
the dominant
contribution comes from the minimum allowed value of $s$ in
the sum in (\ref{eq:fortythree}) which is
$s_m = {p \over {\bar n}}$. For large $p$
the relevant Bessel function in (\ref{eq:fortythree})
behaves as \cite{GRAD}
\begin{equation}
J_{s_m}(\Phi({\bar n},p)) \sim
(2 {p \over {\bar n}}\pi\tanh\beta)^{-(1/2)}
e^{-{p \over {\bar n}}(\beta-\tanh\beta)}
\label{eq:fortyfour}
\end{equation}
where we have defined
\begin{equation}
{\rm cosh}~\beta = (2  |C_{\bar n}| ({2\pi\over L})^2{\bar n}^2 t)^{-1}
\label{eq:fortyfive}
\end{equation}
So long as $\beta > 0$, $G(p)$ is thus exponentially suppressed
for large $p$. The exponential suppression disappears when
$\beta = 0$ or when
\begin{equation}
t = t_0 = (2  |C_{\bar n}| ({2\pi\over L})^2{\bar n}^2 )^{-1}
\label{eq:fortysix}
\end{equation}
The time $t_0$ is {\em exactly} equal to the time of fold
formation $t_f$ as calculated in equation (\ref{eq:twenty}).

When $t > t_f$ the relevant Bessel function behaves, for large
$p/\bar n$ as
\begin{equation}
J_{{p \over {\bar n}}}
(x)~\sim~({2\over {p \over {\bar n}}
\pi\tan\beta})^{(1/2)} \cos(
{p \over {\bar n}} \tan\beta -n\beta -\pi/4)
\label{eq:fortyseven}
\end{equation}
where $\cos \beta = (2  |C_{\bar n}| ({2\pi\over L})^
2{\bar n}^2 t)^{-1}$. At late times
it may be shown that for $p/\bar n>>1$ (but $p << p_M$)
\begin{equation}
G(p) \sim  2  |C_{\bar n}|({2\over L})^2\pi{\bar n}^2 p t~=~p~n_{\rm fold}
\label{eq:fortyeight}
\end{equation}
where
$ n_{{\rm fold}} =
2  |C_{\bar n}|({2\over L})^2\pi{\bar n}^2  t=2\vert p_{max} (x) \vert
{{\bar n} t \over L}$ is the number of folds computed from
the classical motion
of the fermi fluid.

Finally we estimate the quantities $w_n(x,t)$ and see
whether they are nonzero in the classical limit. We will
consider the quantity
\begin{eqnarray}
w_{1,0} (t) & \equiv &\int dx w_1 (x,t) =(2\pi\hbar)^2
\sum_p<\psi(t)|:\alpha_{-p}\alpha_p :|\psi(t)>_c \nonumber \\
& = &  2(2\pi\hbar)^2 \sum_{p>0} G(p,t)
\label{eq:fortynine}
\end{eqnarray}
Since $G(p,t)$ decays exponentially for $p > p_M$, we can
effectively put an upper bound on the sum over $p$ at $p_M$.
\begin{enumerate}

\item For $t < t_f$, one has a $G(p)$ which decays
exponentially with $p$ at rate {\em independent of} $\hbar$
(see equation (\ref{eq:fortyfour})). Thus in this situation
one has $w_{1,0} (t) \sim \hbar^2$ which vanishes in
the classical limit.

\item For $t > t_f$ one has $G(p) \sim p$. In this case one
clearly has $w_{1,0}(t) \sim \hbar^2 p_M^2$. Using (\ref{eq:fournine})
one then has $w_{1,0}(t) \sim O(1)$ and survives in the classical
limit.

\end{enumerate}
Thus we see that
the presence of folds in the classical description signifies
quantum fluctuations of the bosonic field which {\em survive
in the} $\hbar \rightarrow  0$ {\em limit.}

While we have demonstrated our result in a simple model, it
is clear from the derivation that our main contention is valid
for the matrix model described in terms of fermions in an
inverted harmonic oscillator potential  - though the details
would be  more complicated.
 Since  at the quantum level it is sufficient
to have one scalar field to describe states, we have a natural
way to extract the space-time of the string theory from that
of the matrix model.
 The idea is to express an asymptotic state
in terms of the bosonic oscillators $\alpha(\omega)$
and use a quantum map of
the oscillators into oscillators of the string theory tachyon,
$S(\omega)$.
This map is the same as given in \cite{POLTRANS}
\begin{equation}
\alpha(\omega) = {\Gamma (i\omega) \over \Gamma (-i\omega)}~
S(\omega)
\label{eq:fifty}
\end{equation}
States which come out folded at $t = +\infty$ are distinguished
from non-folded states by their lack of coherence.

To summarize, we have shown that the classical limit of the
two dimensional string theory defined as the double scaling
limit of matrix quantum mechanics
is rather nontrivial and differs from that in most
other theories. The number of degrees of freedom needed for
a classical description is more than the number in the
exact quantum theory - the extra degrees of freedom correspond
to nonzero fluctuations {\em even in the} $\hbar \rightarrow
0$ limit. This
feature could not have been guessed from the the low energy
effective action which is {\em locally} Lorentz invariant
and in which unfolded configurations can never develop into
folds; thus any `classical' field configuration  retains its small
($O(\hbar)$) dispersion throughout the evolution.

The fact that a fold corresponds to incoherent quanta agrees with the
spirit of \cite{POLTRANS} where a fold state was termed
`radiation'. The spectrum we find however is not thermal.  In fact
from the discussion above we can estimate the distribution of energies
if, say, a single fold forms when an incident coherent wave reflects
from the tachyon potential in 1+1 string theory.  If the incident wave
has $\sim N^2$ quanta of energy $E$ each, then the state returning
from the wall has, crudely speaking, quanta with energies
$E,~2E,\dots\sim NE$ (with total energy equalling the incident
energy). Thus the mean energy of the returning quanta is higher by a
factor $\sim N$ compared to the incident quanta. In the classical
limit $\hbar\rightarrow 0$, $N\rightarrow\infty$, this frequency
amplification factor goes to infinity.

One possibility is that we cannot trust the naive summation of the
string perturbation theory for non-infinitesimal coupling, and
nonperturbative effects prevent the phenomenon discussed above.
The other possibility is that string theory has a nontrivial sense of
classical limit; in that case one wonders if such phenomena might
occur in higher dimensions as well.
\vskip 2.0cm
\begin{center}
Acknowledgements
\end{center}
We would like to thank A. Jevicki for sharing his insights and for
pointing out \cite{AVJEV}. We would also like to thank M. Li, P. Mende,
 P. Weigmann and B. Zwiebach for discussions. S.R.D would like to
thank the Center for Theoretical Physics at M.I.T., the Physics
Departments of Brown University, Princeton University and Washington
University and the Enrico Fermi Institute for hospitality during the
course of this work. S.D.M. is partially supported by cooperative
agreement number DE-FC02-94ER40818.

\end{document}